\documentclass{osa-article}

\journal{oe}

\articletype{Research Article}
\usepackage{lineno}
\begin{document}

%\title{Experimental signal power evolution shaping in spectral-spatial domains using machine learning-enabled Raman amplifiers}
%\title{Spectral-spatial signal power evolution shaping with experimental design of Raman amplifiers}
\title{Experimental validation of machine-learning based spectral-spatial power evolution shaping using Raman amplifiers}

\author{Mehran Soltani,\authormark{1,*} \href{https://orcid.org/0000-0002-5831-1296}{\includegraphics[scale=0.75]{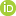}} Francesco Da Ros,\authormark{1} \href{http://orcid.org/0000-0002-9068-9125}{\includegraphics[scale=0.75]{ORCIDiD_icon16x16.png}}, Andrea Carena \authormark{2}\href{http://orcid.org/0000-0001-6848-3326}{\includegraphics[scale=0.75]{ORCIDiD_icon16x16.png}} and Darko Zibar \authormark{1} \href{http://orcid.org/0000-0003-4182-7488}{\includegraphics[scale=0.75]{ORCIDiD_icon16x16.png}}}

\address{\authormark{1}Department of Electrical and Photonics Engineering, Technical University of Denmark (DTU), DK-2800 Kgs. Lyngby, Denmark\\
\authormark{2}Dipartimento di Elettronica e Telecomunicazioni (DET), Politecnico di Torino, Corso Duca degli Abruzzi, 24 - 10129, Torino, Italy\\}

\email{\authormark{*}msolt@dtu.dk} %% email address is required

% \homepage{http:...} %% author's URL, if desired

%%%%%%%%%%%%%%%%%%% abstract %%%%%%%%%%%%%%%%
%% [use \begin{abstract*}...\end{abstract*} if exempt from copyright]

\begin{abstract}We experimentally validate a real-time machine learning framework, capable of controlling the pump power values of Raman amplifiers to shape the signal power evolution in two-dimensions (2D): frequency and fiber distance. In our setup, power values of four first-order counter-propagating pumps are optimized to achieve the desired 2D power profile. The pump power optimization framework includes a convolutional neural network (CNN) followed by differential evolution (DE) technique, applied online to the amplifier setup to automatically achieve the target 2D power profiles. The results on achievable 2D profiles show that the framework is able to guarantee very low maximum absolute error (MAE) (<0.5 dB) between the obtained and the target 2D profiles. Moreover, the framework is tested in a multi-objective design scenario where the goal is to achieve the 2D profiles with flat gain levels at the end of the span, jointly with minimum spectral excursion over the entire fiber length. In this case, the experimental results assert that for 2D profiles with the target flat gain levels, the DE obtains less than 1 dB maximum gain deviation, when the setup is not physically limited in the pump power values. The simulation results also prove that with enough pump power available, better gain deviation (less than 0.6 dB) for higher target gain levels is achievable.
\end{abstract}

%%%%%%%%%%%%%%%%%%%%%%%%%%  body  %%%%%%%%%%%%%%%%%%%%%%%%%%
\section{Introduction}

Distributed Raman amplifiers (DRAs) present several advantages over the erbium doped fiber amplifiers (EDFAs) including lower noise figure (NF), and flexibility in power evolution design by adjusting the pump power and wavelength values\cite{Agrawal, 7163281}. Due to multi-pumping scheme, DRAs are also a practical solution to amplify a broad range of wavelengths beyond the C-band and increase the available transmission capacity\cite{deMoura:20}. One approach in designing Raman amplifiers is to optimize the pump power and wavelength values to obtain a desired signal power evolution shape, jointly in spectral and spatial (fiber distance) domains. Shaping the signal power evolution in frequency and distance is a beneficial way to approach some of the long-term goals in optical communication systems such as signal-to-noise (SNR) maximization and nonlinearity mitigation \cite{PhysRevLett.101.123903, tan2018distributed}. As an example, a flat two-dimensional (2D) profile in frequency and distance, resembling a lossless link, minimizes the accumulated amplified spontaneous emission (ASE) noise at the end of the fiber \cite{Ania-Castanon:04, 1601058, PhysRevLett.101.123903}. This flat 2D profile is also a requirement for the transmission based on Nonlinear Fourier Transform (NFT) \cite{Mollenauer:88, le2015nonlinear}. Another example is a 2D symmetric power profile with respect to the middle point in distance. A symmetric 2D profile can be utilized to mitigate the nonlinear impairments using optical phase conjugation (OPC) systems \cite{Phillips:14, tan2018distributed, s22030758}.  

Power profiles in a 2D space are mostly addressed by heuristic tuning of the Raman pump power and wavelength values, where the pump power values are adjusted based on experience and without applying an intelligent optimization framework \cite{1561354, PhysRevLett.101.123903, Rosa1515, Bednyakova:13}. The heuristic optimization of the parameters in a practical setup requires simplifying the optimization problem which can be challenging when the dimensionality of the problem to solve, i.e. the number of parameters to optimize, increases. This approach will be time consuming and also less accurate in case the goal is to design several 2D target profiles with different cost functions using the same Raman amplifier setup under test. In \cite{Soltani:21, 9721641}, we presented and numerically validated an online machine learning framework to automatically optimize the Raman pump power values to design power profiles of practical interest, jointly in frequency and fiber distance domains. The proposed approach consists of a convolution neural network (CNN) \cite{Soltani:21}, trained as an inverse system model, followed by a differential evolution (DE) \cite{9721641} technique. Numerical results shown that the resulting framework has the flexibility to design different 2D profiles with specific objectives by optimizing the pump power values.

In \cite{soltani22}, we presented a Raman amplifier setup and experimentally verified the performance of the CNN model and the CNN-assisted DE framework to design a data-set of achievable 2D power profiles. This amplifier setup utilizes four first order counter-propagating pumps and the target data-set is generated by applying random pump power values to the setup and measuring their resulting 2D power profile. For each target 2D profile in the data-set, the pump power values are optimized aiming to minimize the maximum absolute error (MAE) between the target and designed 2D profile. First, the CNN prediction on the target 2D profiles is investigated, which results in low MAE on average (0.37 dB), while showing high MAE($>$1 dB) for roughly 2\% of the profiles. To improve the CNN accuracy on 2D profiles with high MAE, the CNN-assisted DE is applied to perform online optimization on the Raman amplifier setup, resulting in less than 0.5 dB for all of the selected high-error 2D profiles. 

In this paper, we extend \cite{soltani22} by providing more details on the experimental validation of the CNN model and the CNN-assisted DE framework for designing achievable 2D profiles. Additionally, in this work we experimentally test the DE framework in a scenario where the target is to shape the 2D signal power evolution to jointly satisfy multiple spatial-spectral objectives, rather than designing a specific achievable 2D profile shape. In particular, we consider two cost functions as the objectives and perform an online optimization of the pump power values to minimize these cost functions using the Raman amplifier setup. The two cost functions are: 1) the maximum deviation from a spectrally flat-gain profile at the end of the fiber (over the full C-band), and 2) the maximum spectral power excursion along the fiber distance. This is a multi-objective optimization problem with non-differentiable cost functions with respect to the free parameters, which is challenging to be solved with gradient-based neural network (NN) approaches presented in \cite{9252846}. Our approach in this scenario is to use only the DE framework without involving the CNN. The proposed DE framework provides less than 1 dB gain deviation and spectral power excursion when enough pump power value is provided in the amplifier setup. Additionally, the simulation results also assert that the DE is able to achieve less than 0.6 dB gain deviation when the amplifier setup is not limited by the upper-bound pump power values.  

The paper is organized as follows. Section 2 presents the experimental Raman amplifier setup used to validate the proposed machine learning-based power optimization framework. In Section 3, the proposed framework is discussed in more details with particular emphasis to the online application. Section 4 presents and compares the experimental results of the DE and the CNN-assisted DE frameworks. Finally, section 5 concludes the paper.

\section{Experimental setup}

\begin{figure}[htb!]
\centering\includegraphics[width=14cm]{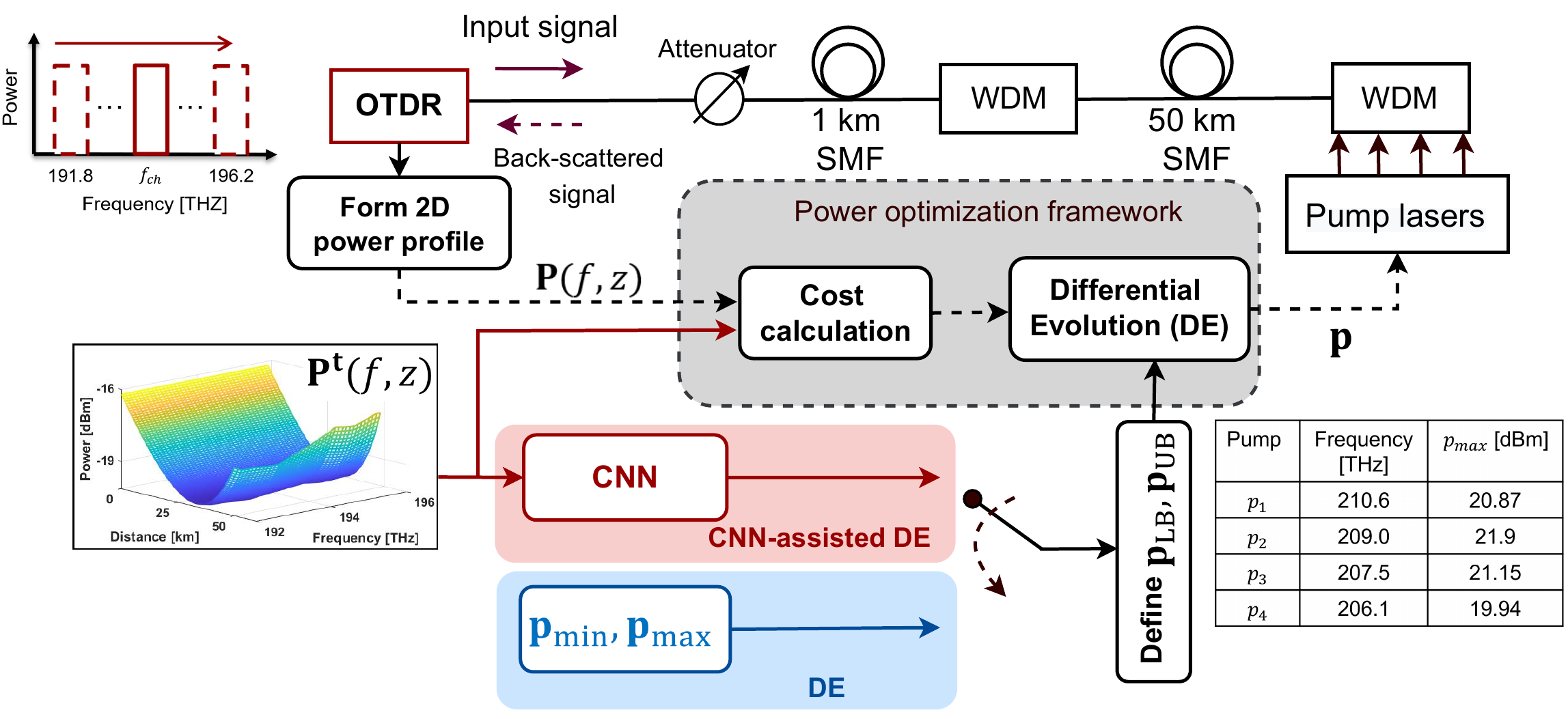}
\caption{The experimental setup and the block diagram of the processes used to optimize the pump powers values for designing a 2D target power profile $\textbf{P}^\textbf{t}(f,z)$.}
\label{exp_setup}
\end{figure}

The experimental setup for verifying the proposed machine learning-based framework to optimize the Raman pump power values is depicted in Fig.\ref{exp_setup}. We consider a span of standard single-mode fiber (SSMF) with 50 km length and the Raman pump module of four counter-propagating lasers. Pump wavelengths are fixed (shown with their maximum available power value $p_{max}$ in the table inset in Fig. \ref{exp_setup}), and able to amplify the entire C-band. The goal is to optimize the Raman pump powers values  $\textbf{p} = [p_1, p_2, p_3, p_4]$ to achieve the targeted 2D profile  $\textbf{P}^\textbf{t}(f,z)$, defined in both spectral ($f$) and spatial ($z$) domains. 

To measure the power evolution over the spectrum and along the fiber distance, a frequency-tunable optical time-domain reflectometer (OTDR) is used. The signal bandwidth of the OTDR covers the C-band between 191.8 THz and 196.2 THz, equivalent to 44 channels with 100 GHz spacing. The OTDR is connected to the fiber span with three elements in between: a tunable attenuator employed to control the signal power flow into the OTDR, a 1 km SMF link to cover the dead-zone of the OTDR, and a wavelength division multiplexer (WDM), utilized to filter out the pump waves in the range between 203.9 THz and 211.1 THz, such that they do not enter the OTDR. Another WDM coupler is placed at the end of the fiber link to combine the signal and the pumps. 

The OTDR introduces a low power signal (-12 dBm) into each of the channels and measures their back-scattered signal, sequentially. Overall attenuator and WDM loss is 4 dB resulting in almost -16 dBm input signal power to the 50 km fiber. To reduce the impact of noise on the measurements and have a clear trace for each channel, the signal pulse width is set to 3 $\mu s$ and the distance resolution of the OTDR to 8.2 m. The OTDR measures the signal power evolution for all channels, and we post-process the measured signal by applying a Savitzky-Golay smoothing filter \cite{ac60214a047} with window size $w=19$ (equivalent to 19$\times$8.2=155.8 m) and polynomial order $n=2$ in distance to reduce the signal fluctuations. According to the numerical results reported in \cite{Soltani:21, 9721641}, a distance resolution of 500 m is sufficient for the CNN training and evaluation. Therefore, the smoothed traces are down-sampled in distance direction using a linear interpolation to achieve 500 m resolution, and a 2D power profile $\textbf{P}(f,z)$ of size $44 \times 100$ is formed. $\textbf{P}(f,z)$ is used as the input to the cost calculation block, where its MAE value with respect to the pre-defined target 2D profile $\textbf{P}^\textbf{t}(f,z)$ is calculated. After the cost calculation, the power optimization framework updates the pump powers accordingly and applies a new set of pump powers to the setup, aiming to reduce the MAE in the following iteration (see Section 3 for the framework description). The process of applying pump powers, recording the resulting 2D profile, and the cost calculation continues until a convergence criteria such as a minimum cost value without considerable variation in pump power values or a maximum number of DE iterations is achieved. 

\section{Raman pump power optimization framework}

The proposed pump power optimization framework, shown in Fig.\ref{exp_setup}, consists of an evolutionary optimization algorithm, known as differential evolution (DE) \cite{app8101945}. The main goal of the DE is to optimize the pump power values to achieve the target 2D power evolution profile $\textbf{P}^\textbf{t}(f,z)$ in the span. This framework is evaluated numerically in \cite{9721641} using the synthetic data to find the optimum pump powers values for 2D profiles of practical interest. The DE can be used to perform optimization in an online closed-loop process by applying pump power values to the setup, evaluating their resulting cost, and updating them aiming to reduce the cost in the next iteration. If the DE optimization starts with a randomly selected population of individuals (as the initial candidate solutions), it will be more time consuming and also prone to local minimum in case the individuals in the initial population are not in the vicinity of the solution. Additionally, if the number of parameters to be optimized is high, the volume of the space to explore increases, and consequently, the DE can converge to different local minimum points each time it starts to run. For a target 2D profile $\textbf{P}^\textbf{t}(f,z)$, it is shown in \cite{9721641} that a CNN-based inverse model initialization improves the DE performance in terms of its finally achieved cost value, the speed of the convergence and also the certainty in converging to the same optimum point every time the DE starts to run. The CNN model, which consists of four CNN layers followed by two fully-connected hidden layers, is trained to learn the mapping between the 2D profiles and their corresponding pump powers values using a data-set. The data-set is generated by applying randomly selected pump powers values to the amplifier setup and measuring their resulting 2D power evolution profile. 
% Numerical results have shown that on average, the CNN achieves low MAE on the test data-set, containing the 2D power profiles which were unseen by the model during the training. However, due to off-line training, it is not capable of further improvement if it is not highly accurate on a test 2D profile. In spite of this, the CNN prediction can be used as an initial solution for a real-time optimization scheme such as DE. 

To experimentally validate the pump power optimization framework, two DE population initialization approaches are proposed as shown in Fig.\ref{exp_setup}. In the first approach, so-called CNN-assisted DE in Fig.\ref{exp_setup}, the target 2D profile $\textbf{P}^\textbf{t}(f,z)$ is used as the input to CNN model, where a set of pump powers values $\textbf{p}'$ are predicted as the initial solution. These values are then used to define the pump power lower-bound and upper-bound values $\textbf{p}_{LB}$ and $\textbf{p}_{UB}$, respectively. The $\textbf{p}_{LB}$ and $\textbf{p}_{UB}$ are specified according to the deviation of the first-order pump values in \cite{9721641} by considering $\textbf{p}_{LB}=\textbf{p}'- 0.5\cdot \textbf{p}'$ and $\textbf{p}_{UB}=\textbf{p}'+ 0.5\cdot \textbf{p}'$. 
In the second experimental approach, we have proposed the DE with random initialization, so-called DE in Fig.\ref{exp_setup}. In this approach, the CNN is not involved and each element of $\textbf{p}_{LB}$ and $\textbf{p}_{UB}$ are defined without any prior knowledge and basically are the minimum $p_{min} = -5$ dBm and maximum $p_{max}$ pump powers values (provided in the table inset in Fig. \ref{exp_setup}), respectively. To provide intuition on how the CNN improves the DE initialization, the performance of the DE and the CNN-assisted DE is compared in terms of their MAE in designing a data-set of achievable target 2D profiles in the proposed amplifier setup.

\section{Experimental results}

\subsection{Designing achievable 2D profiles}In this section, we investigate and compare the performance of the CNN-assisted DE and the DE on predicting the pump power values for a test data-set of achievable target 2D profiles. This data-set consists of 500 set of pump powers and their resulting measured 2D profiles. For each test 2D profile the goal is to predict the pump power values aiming to minimize the error between the true and the predicted pump power values, or the MAE between the target 2D profile and the resulting 2D profile in the span. Both CNN-based and random initialization approaches are used and compared in terms of convergence speed and resulting MAE. We start with the validation of the CNN-assisted DE framework, which its first step is to train and evaluate the CNN model. Regarding this, 4400 2D profiles are generated in total, divided into two sets for train and validation with the size of 4100 and 300 2D profiles, respectively. 
% It should be noted that the all test 2D profiles and their pump power values are compared to the data in both train and validation sets to avoid overlap and keep the test data unseen during the training and validation of the CNN. 
% The training data-set has 4100 number of 2D profiles with their corresponding true pump powers values, while the actual training size is specified based on the prediction performance on the validation data. To clarify the best training size, we create training subsets of size varying between 1500 and 4100 by random selection of the profiles from the main training set (including 4100 2D profiles), and use them to train the CNN. At each training size, the CNN is trained using the training subset and evaluated based on the prediction performance on the validation set. According to these analyses, a subset with 3700 2D profiles is selected as the final training size.

Once the CNN is trained, the test 2D profiles are used as the input to the model and the $R^2$ score is calculated to evaluate the correlation between the true and the predicted pump power values. The $R^2$ score takes the values between 0 and 1 where the highest value indicates a perfect prediction. The $R^2$ score attained for each pump is reported in Table \ref{r values}. According to this results, in general, the CNN has good performance on predicting all four pump power values with more accurate prediction for $p_3$ and $p_4$, compared to the $p_1$ and $p_2$. The lower accuracy of $p_1$ and $p_2$ (compared to $p_3$ and $p_4$) is mainly because of their less impact on the signal power evolution shape as the peak of their corresponding Raman gain efficiency lies slightly outside the signal bandwidth (between 191.8 THz and 196.2 THz). 

\begin{table}[h]
\caption{$R^2$ test scores for the CNN model prediction.}
\label{r values}
\centering
\begin{tabular}{ | m{1.75cm} | m{0.9cm}| m{0.9cm} | m{0.9cm}| m{0.9cm} | } 
     \hline
    Pump  & $p_1$ & $p_2$ & $p_3$ & $p_4$ \\ \hline
    $R^2$ & 0.86 & 0.87 & 0.91 & 0.93\\ [3pt]
     \hline
\end{tabular}
\end{table}

\begin{figure}[hbt!]
\centering\includegraphics[width=10cm]{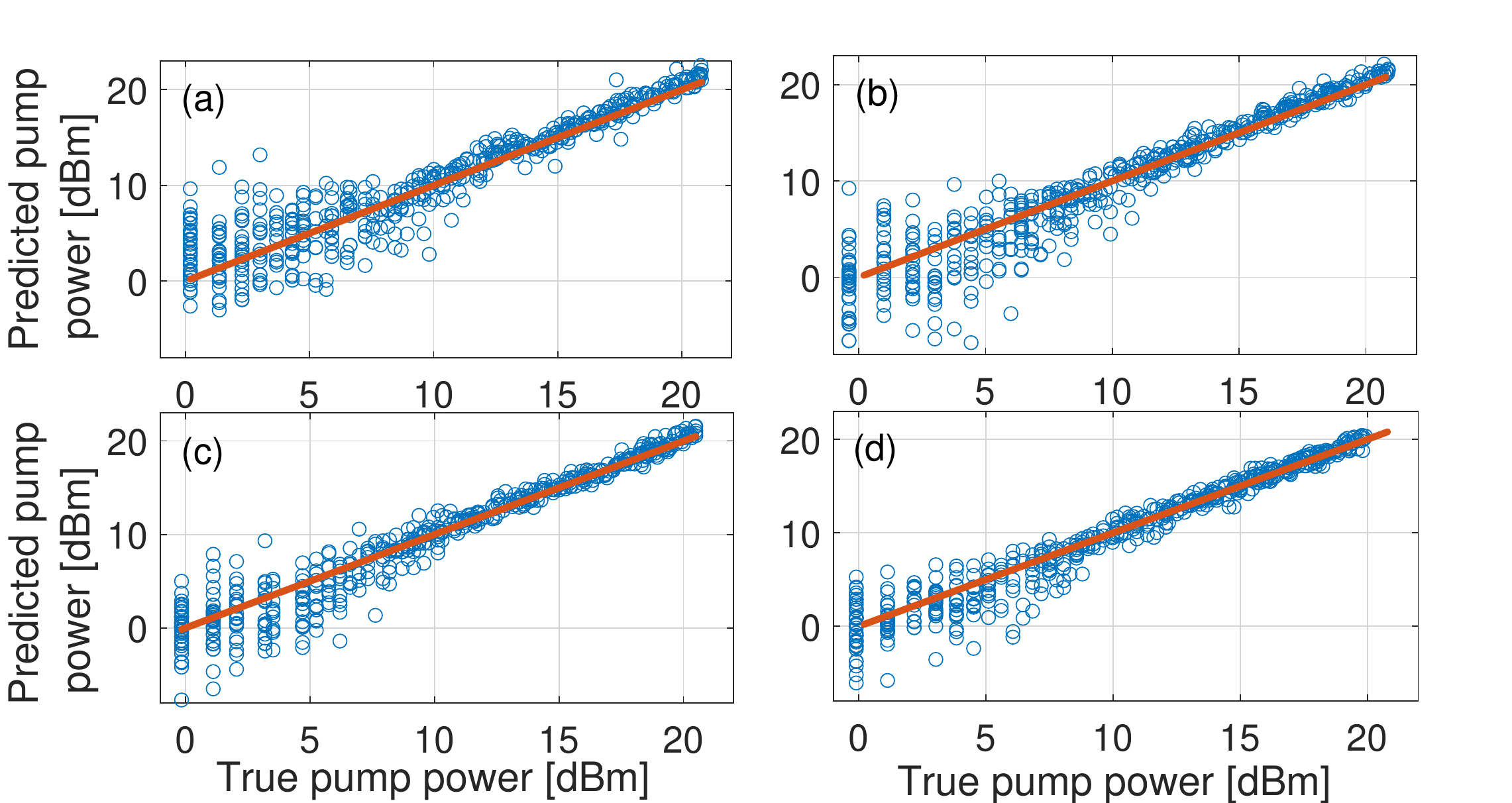}
\caption{The scatter plot of the true versus predicted pump power values using the CNN model on test data. Each blue dot corresponds to a test 2D profile and the orange solid line represents the ideal prediction. (a) $p_1$, (b) $p_2$, (c) $p_3$, (d) $p_4$. }
\label{CNN_true_predicted}
\end{figure}

Additionally, a scatter plot of the true and the predicted pump power values is shown in Fig \ref{CNN_true_predicted} (a)-(d) for all four pump values $p_1$, $p_2$, $p_3$, $p_4$, respectively. Based on this figure, the correlation between the true and the predicted pump power values is low for low ranges of pump power values which is because of the low impact of the low pump power values on the signal power evolution shape. As the pump power values increase, they become more influential on the signal power profile, and the prediction accuracy increases, accordingly.

\begin{figure}[h!]
\centering\includegraphics[width=13cm]{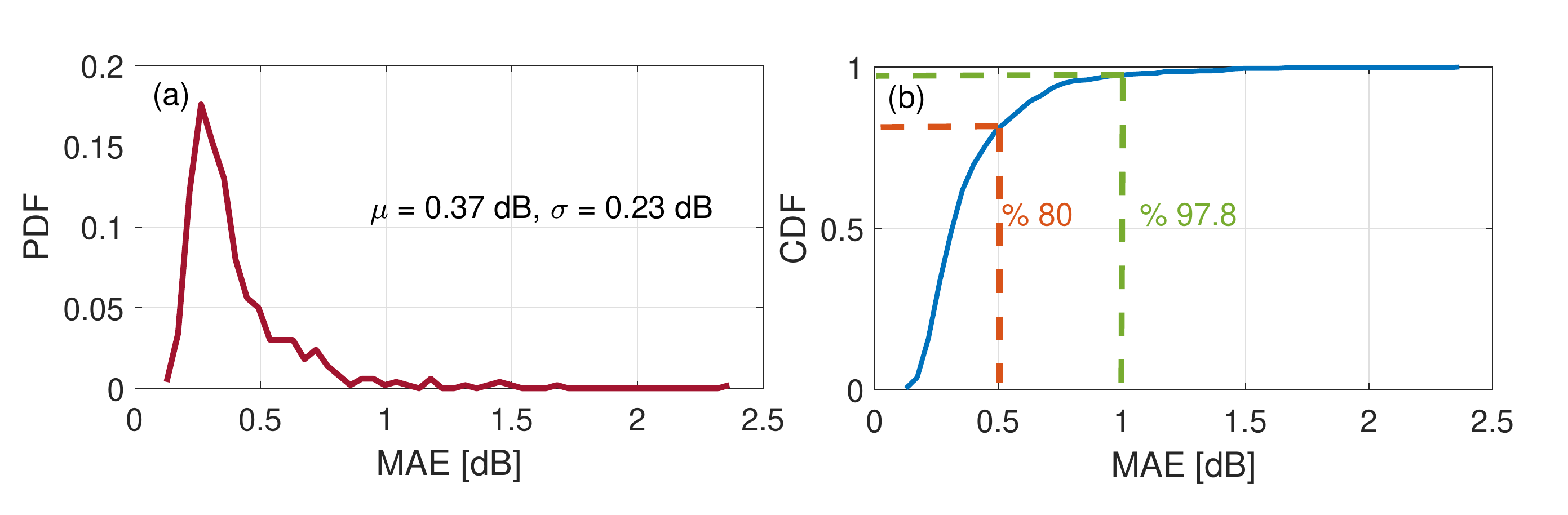}
\caption{The CNN model performance on test profiles. (a) PDF of the MAE, (b) CDF of the MAE.}
\label{CNN_pdf}
\end{figure}

A more rigorous and meaningful approach to evaluate the CNN prediction performance is to apply the predicted pump power values to the experimental amplifier setup, measure the designed 2D profile and calculate the MAE between the targeted 2D profile and the resulting one. Fig.\ref{CNN_pdf} (a) and (b) show the probability density function (PDF) and the cumulative density function (CDF) of the MAE for all test profiles, respectively. For the test 2D profiles, the CNN achieves the MAE with mean $\mu = 0.37$ dB, and standard deviation $\sigma = 0.23$ dB. According to the CDF results, 80\% of the test 2D profiles have MAE lower than 0.5 dB, and 97.8\% of them result in MAE values lower than 1 dB. 

According to Fig. \ref{CNN_pdf}, the CNN obtains statistically low MAE values, while only for eleven 2D profiles ($\approx$ 2.2\% of all 500 test 2D profiles), it results in a MAE higher than 1 dB. Selecting this 2D profiles with the error higher than 1 dB, we apply the CNN-assisted DE approach to improve the prediction results in real-time on the experimental setup with a limit of maximum 100 iterations.

\begin{figure}[h!]
\centering\includegraphics[width=13cm]{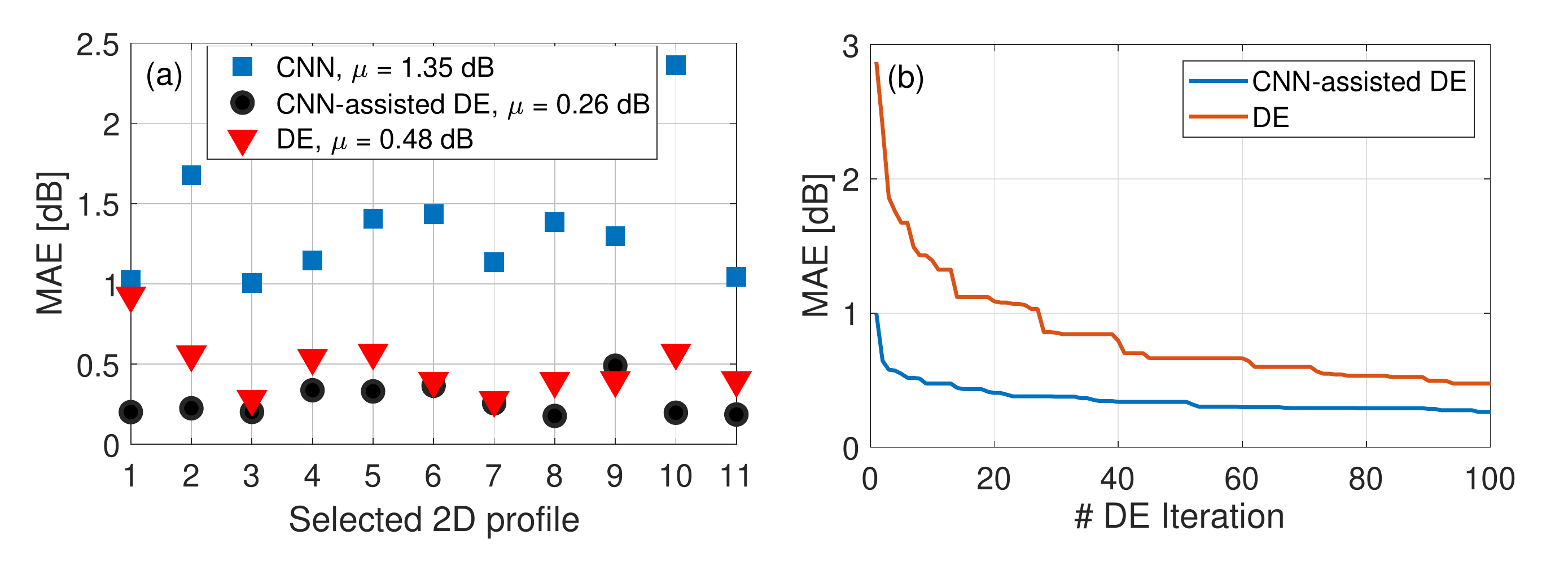}
\caption{(a) The MAE for the CNN model, the CNN-assisted DE and the DE approaches for the eleven selected 2D profiles with high MAE. (b) The average value of the MAE over the selected 2D profiles for each DE iteration using the CNN-assisted DE and the DE scenarios.}
\label{Subplot_Error_versus_iteration_10_sample}
\end{figure}

\begin{figure}[h!]
\centering\includegraphics[width=12.5cm]{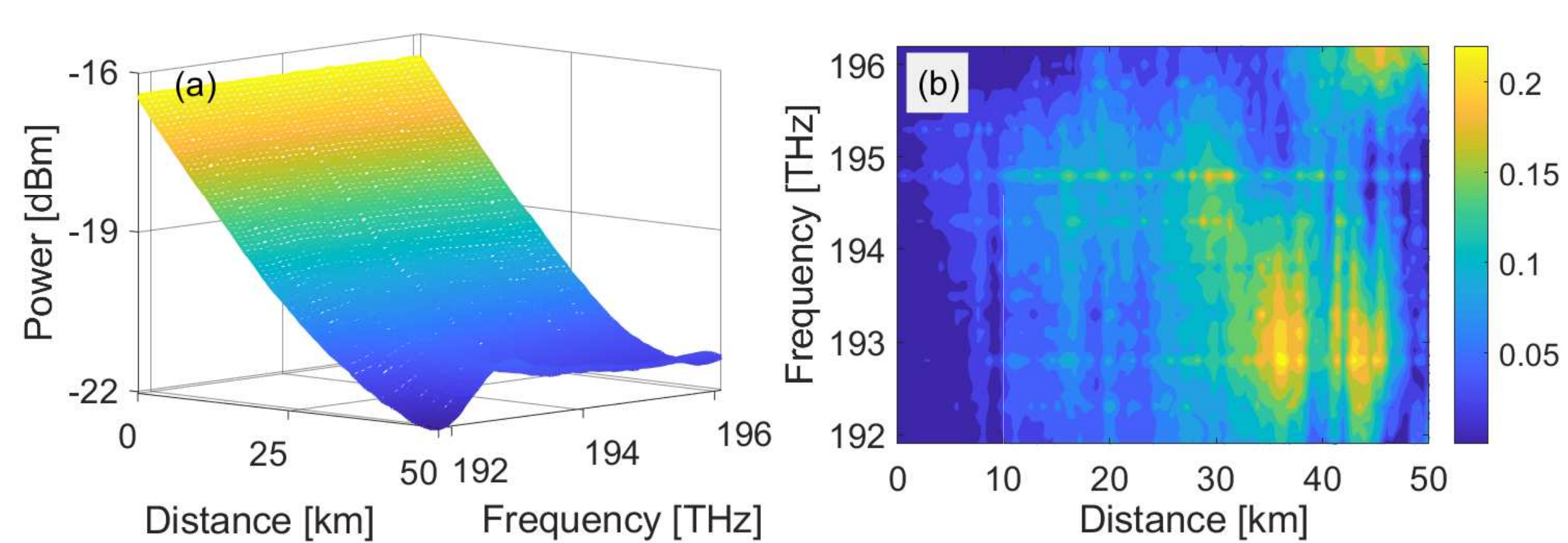}
\caption{CNN-assisted DE result for the 10th selected 2D profile in Fig.\ref{Subplot_Error_versus_iteration_10_sample} (a).(a) Target 2D profile, (b) Heatmap of the absolute error (in dB) between the target and the predicted 2D profiles over the frequency and distance domains.}
\label{True_predicted_sample_489_v1}
\end{figure}

To provide an intuition regarding the CNN impact on the DE initialization, another set of experiments for the proposed eleven 2D profiles is performed using  DE. For each one of these selected 2D profiles, indexed from 1 to 11, the MAE is shown in Fig \ref{Subplot_Error_versus_iteration_10_sample} (a) for the CNN, the CNN-assisted DE and the DE scenarios. The resulting error using CNN-assisted DE for all eleven 2D profiles is less than 0.5 dB. Fig \ref{Subplot_Error_versus_iteration_10_sample} (a) also asserts that the CNN-assisted DE results are considerably better than the CNN and the DE results. In addition, the average MAE evolution for all eleven 2D profiles over the number of DE iterations is shown in Fig.\ref{Subplot_Error_versus_iteration_10_sample} (b) for the CNN-assisted DE and DE scenarios. This plot shows that the DE initialized with the CNN converges faster and to a point with lower MAE.

Moreover, a visual representation of the CNN-assisted DE framework result (10th selected 2D profile in Fig.\ref{Subplot_Error_versus_iteration_10_sample}) is shown in Fig.\ref{True_predicted_sample_489_v1}, providing the target 2D profile Fig. \ref{True_predicted_sample_489_v1} (a) and the resulting heatmap of its absolute error with the predicted 2D profile over the frequency and fiber distance Fig. \ref{True_predicted_sample_489_v1} (b). For this case, the resulting MAE value between the target and the predicted 2D profile is 0.22 dB.

\subsection{Designing objective-based 2D profiles}

In this section we approach a different design scenario where the target is not to minimize the MAE between a target 2D profile and the designed 2D profile, but to achieve a 2D profile which fulfills multiple desired spectral-spatial objectives. In the following scenario, the specific objective is designing 2D profiles with a desired flat gain level at the end of the span, meanwhile, minimizing the spectral excursion in the entire span. This is a multi-objective optimization problem which is quite complex to solve, and we experimentally prove that it can be approached with the DE, applied real-time to the amplifier setup. It is worth noting that the CNN-assisted in Fig.\ref{exp_setup} is not applicable since there is no 2D target profile $\textbf{P}^\textbf{t}(f,z)$ to be used as the input to the CNN model. Only the DE can be used effectively for this design scenario, therefore, we define two cost functions for this purpose. The first objective is to have a spectrally flat 2D power evolution in distance. For this objective, the first cost function $J_0(\textbf{p})$, referred to as the maximum spectral power excursion, which is aimed to be minimized, is formulated as the following:

% In this section we experimentally evaluate a designing scheme in which the target is to minimize an arbitrary cost function defined based on a spectral-spatial property, rather than minimizing the MAE between a 2D target profile and the resulting one. Similar to some of the design approaches in the literature, a 2D target profile is not of the interest here, but a minimum cost function on the resulting 2D profile such as the minimum spectral excursion or a specific gain level at the receiver side is approached. Considering this approach, the scheme (1) cannot be used since there is not a 2D target profile $\textbf{P}^\textbf{t}(f,z)$ to be used as the input. While, the scheme (2) can be used effectively for this design problem. 

%In this section, we experimentally evaluate a designing scheme in which the target is not to design a test 2D profile, but to minimize a cost function on the resulting 2D profile in the span. This cost function is defined based on a spectral-spatial property such as the minimum spectral excursion or a target gain level at the receiver side. Considering this goal,

\begin{equation}
   J_0(\textbf{p}) =\\ \smash{\displaystyle\max_{z}[\smash{\displaystyle\max_{f}(\textbf{P}(f,z|\textbf{p}))} - \smash{\displaystyle\min_{f}(\textbf{P}(f,z|\textbf{p}))}]}
\label{eq:F1}
\end{equation}

\begin{figure}[htb!]
\centering\includegraphics[width=12.5cm]{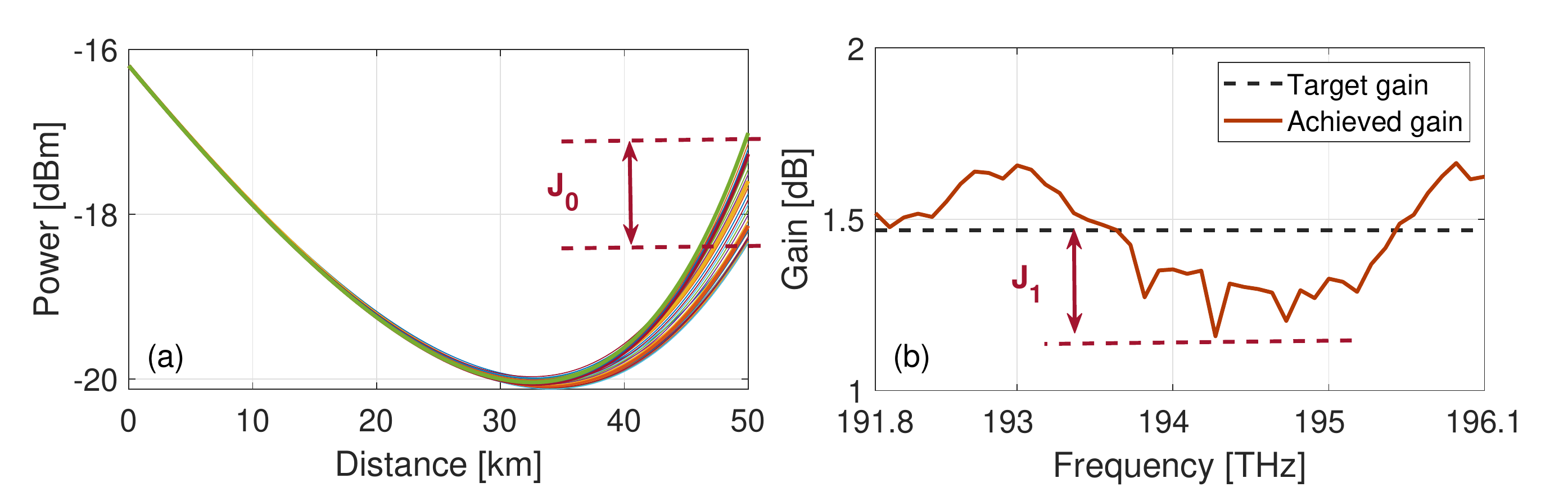}
\caption{(a) Spatial representation of $J_0$, (b) Spectral representation of $J_1$.}
\label{power_2D_j_0_j_1_subplot}
\end{figure}

Minimizing $J_0(\textbf{p})$ results in achieving a 2D profile which has the minimum spectral excursion over the entire span. To provide a visual intuition on $J_0$ value, Fig.\ref{power_2D_j_0_j_1_subplot} (a) shows a sample power evolution profile for all channels over the distance, and the the maximum spectral excursion as $J_0$ is specified, which for this specific case, occurs at the end of the span.

The second objective is to achieve a spectral flat target gain level such as $\textbf{g}^t(f)$ at the end of the span. Approaching this objective, the cost function $J_1(\textbf{p})$ is defined as the maximum absolute deviation between the achieved gain $\hat{\textbf{g}}(f,\textbf{p})$ and the target gain level $\textbf{g}^t(f)$ at the end of the fiber, formulated as:

\begin{equation}
     J_1(\textbf{p}) =  \displaystyle\max_{f}|\smash{\hat{\textbf{g}}(f,\textbf{p})- \textbf{g}^t(f) }|
\label{eq:F2}
\end{equation}

where $L$ is the span length, and the on-off gain $ \hat{\textbf{g}}(f,\textbf{p})$ is defined as:

\begin{equation}
    \hat{\textbf{g}}(f,\textbf{p}) = \textbf{P}(f,z=L|\textbf{p})- \textbf{P}(f,z=L|\textbf{p}_{off})
\end{equation}

where $\textbf{P}(f,z|\textbf{p}_{off})$ is the 2D power profile when all pumps are turned off. In Fig \ref{power_2D_j_0_j_1_subplot} (b), a target gain level is targeted and the $J_1$ value is specified according to a sample 2D profile's achieved gain at the end of the fiber. 

Considering the above-mentioned objectives, we define the multi-objective optimization problem aiming to minimize both $J_0(\textbf{p})$ and $J_1(\textbf{p})$, simultaneously, and find the optimal set of pump power values $\textbf{p}^*$. To make the optimization process simpler and also to be able to control the impact of each objective on the final result, we make an approximation by converting the multi-objective optimization into a classical weighted-sum as proposed in \cite{9721641}. Approaching this, each objective is multiplied by a weight, defined as a hyper-parameter, and it is added to the other objectives as the following: 

% \begin{align}
% \label{eq:so}
% &\textbf{p}^{*} = \smash{\displaystyle\underset{\textbf{p}}{\arg\min} \sum_{i=0}^{1}{m_iJ_i(\textbf{p})}} 
%  \\
%  &\text{such that   }\textbf{p}_{LB} \leq \textbf{p}_{pump} \leq \textbf{p}_{UB}, \;  m_i>0, \;  \sum_{i=0}^{1}{m_i}=1  \notag \\
%  \notag
% \end{align}

\begin{align}
\label{eq:so}
\centering
&\textbf{p}^{*} = \smash{\displaystyle\underset{\textbf{p}}{\arg\min} [m_0J_0(\textbf{p}) + m_1J_1(\textbf{p})]} 
 \\\notag
 \\
 &\text{such that   }\textbf{p}_{LB} \leq \textbf{p}_{pump} \leq \textbf{p}_{UB}, \;  m_0, m_1>0, \;  m_0+m_1=1  \notag \\
 \notag
\end{align}

\begin{figure}[htb!]
\centering\includegraphics[width=12.5cm]{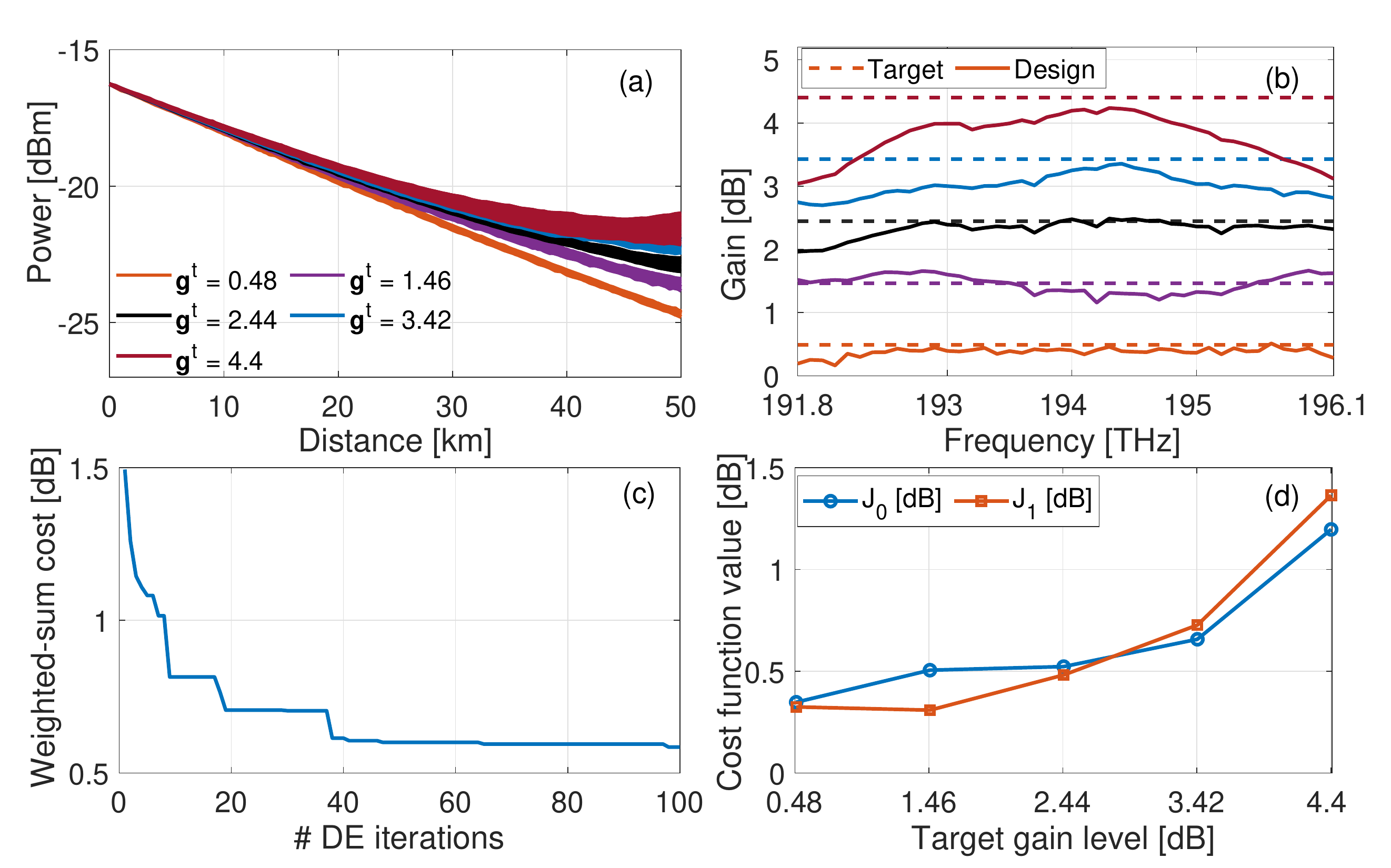}
\caption{DE results of pump power optimization by solving Eq.(\ref{eq:so}) with different target gain levels. (a) Spatial representation of the resulting power evolution profiles over the distance. (b) Spectral representation of target and designed gain levels at span end. (c) The average cost value over the five 2D profiles for each DE iteration. (d) $J_0$ and $J_1$ values achieved for different target gain levels.}
\label{subplot_flat_spectrum_results}
\end{figure}

where $m_0$ and $m_1$ are the weights used to control the impact of $J_0$ and $J_1$ on the optimal point. In our analyses, we set $m_0 = m_1 = 1/2$, giving $J_0$ and $J_1$ the same impact. Five 2D profiles with different flat gain levels are designed by solving the Eq. (\ref{eq:so}). The maximum possible gain level provided by the pumps at the end of the fiber,  when all of the pumps are operating at their maximum available power, is equal to $g_{max} = 4.9$ dB. Considering this, we target five equally-spaced gain levels starting with 0.48 dB and ending with 4.4 dB. 
% Considering this, we have targeted five gain levels equal to $G^t_1 = 0.1G_{max}$, $G^t_2 = 0.3G_{max}$, $G^t_3 = 0.5G_{max}$, $G^t_4 = 0.7G_{max}$ and $G^t_5 = 0.9G_{max}$.
In  Fig.\ref{subplot_flat_spectrum_results} (a), the resulting five power evolution profiles with different flat target gains $\textbf{g}^t$ are shown over the distance, with their corresponding spectral gain versus the target gain level, depicted in Fig.\ref{subplot_flat_spectrum_results} (b). Additionally, the average of the cost value for all five profiles at each DE iteration is calculated and shown in Fig.\ref{subplot_flat_spectrum_results} (c), demonstrating that the average cost does not significantly improve after 40 iterations. 

\begin{figure}
\centering\includegraphics[width=12.5cm]{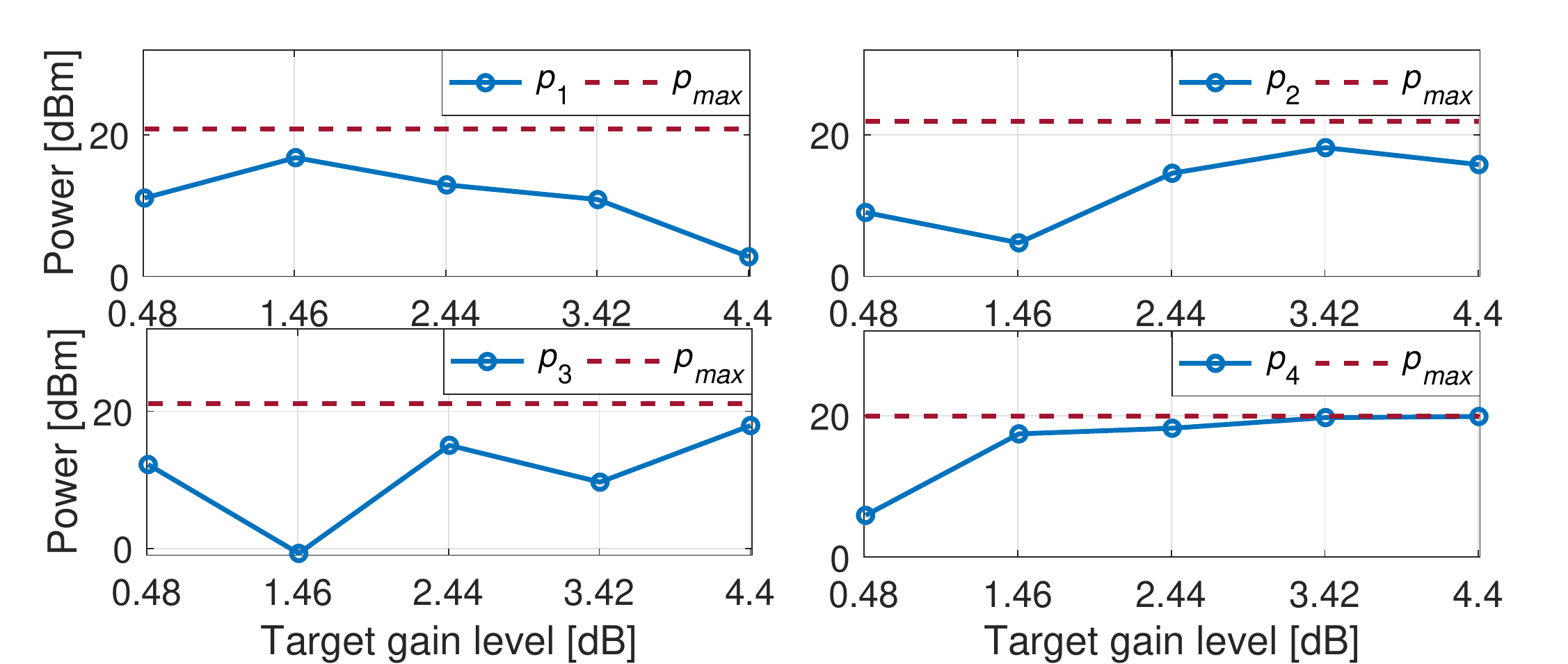}
\caption{Resulting pump power values in the experimental analyses for different target gain levels.}
\label{optim_pump_powers_flat_gains_subplot_experimental}
\end{figure}

Moreover, the cost values of $J_0$ and $J_1$ over the different target gain levels are illustrated in Fig.\ref{subplot_flat_spectrum_results}(d), showing less than 1 dB error for target gain levels up to 4 dB. Both $J_0$ and $J_1$ increase noticeably when the target gain level reaches higher than 4 dB. This phenomenon is mainly because of the limited pump power provided by the pumps in the setup depicted in Fig.\ref{exp_setup}, rather than the failure of the proposed power optimization framework. As the target gain level increases, the experimental amplifier setup physically limits the performance of the DE framework in achieving low cost 2D profiles by providing low upper-bound pump power $\textbf{p}_{UB}$ values. Regarding this, we have shown the resulting power values of all four pumps for different target gain levels by applying the DE framework in Fig.\ref{optim_pump_powers_flat_gains_subplot_experimental}. In particular, it is shown that the power value of pump $p_4$, with increasing the target gain level, has an increasing trend towards its maximum value 19.94 dBm. In case the pumps, especially $p_4$, provide enough power value and they do not limit the optimization process, the DE would be able to design target gain levels more accurately.

\begin{figure}[h]
\centering\includegraphics[width=12.5cm]{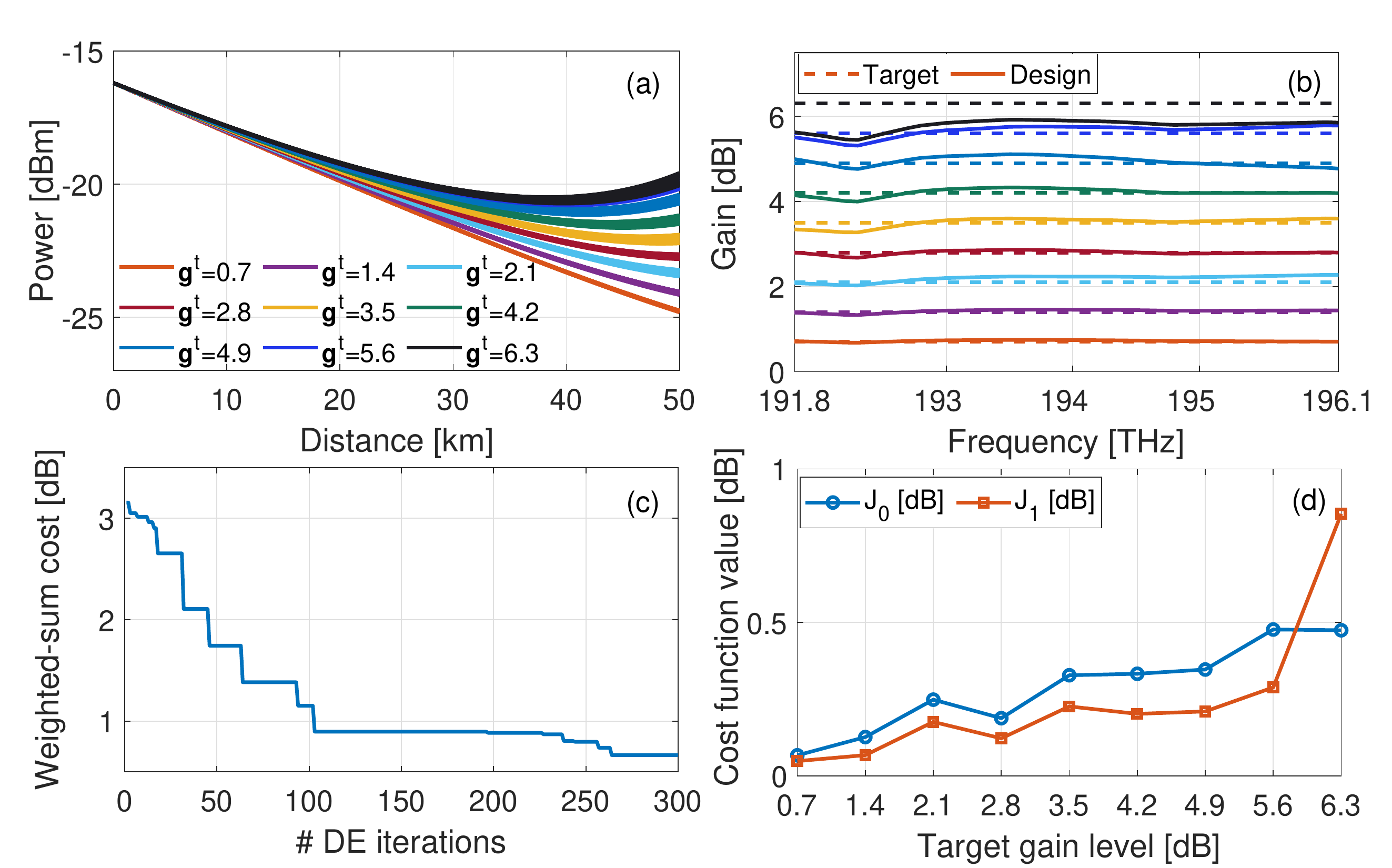}
\caption{Simulation results of pump power optimization by solving Eq.(\ref{eq:so}) with different target gain levels. (a) Spatial representation of the resulting power evolution profiles over the distance. (b) Spectral representation of target and designed gain levels at span end. (c) Average cost value over the five 2D profiles for each DE iteration. (d) $J_0$ and $J_1$ values achieved for different target gain levels.}
\label{subplot_simulation_flat_v1}
\end{figure}

\begin{figure}[h!]
\centering\includegraphics[width=12.5cm]{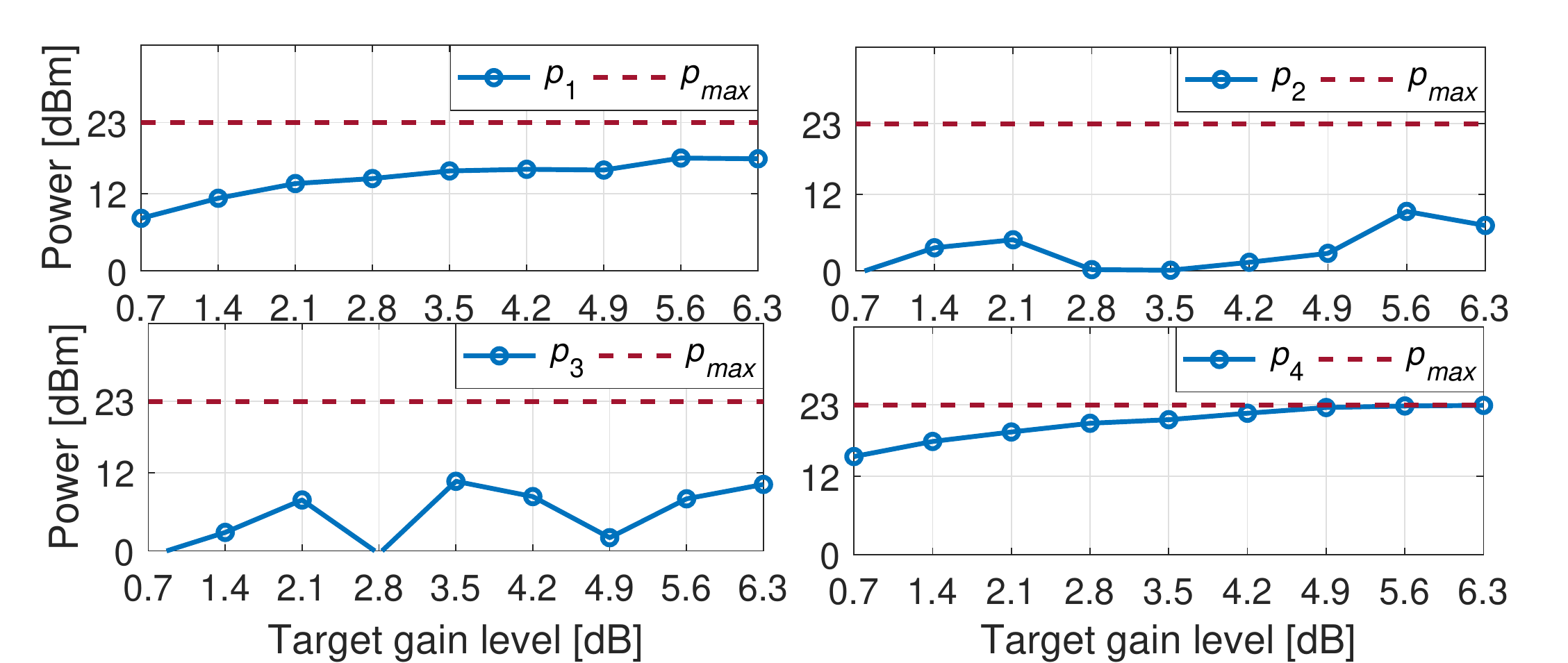}
\caption{Resulting pump power values in the simulation analyses for different target gain levels.}
\label{Optim_pump_power_values_subplot_simulation}
\end{figure}

To confirm this statement, we have performed a set of simulations to analyze the amplifier setup in Fig.\ref{exp_setup} with the same configuration, except that each pump in simulations provides a higher power value, up to 23 dBm. In addition, since the range of space to explore with the new pump power ranges has higher volume compared to the experimental analyses, we set the number of the DE iterations to 300 (higher than 100 iterations in the experimental analyses).

In the simulation analyses, we use the same weight values( $m_0 = m_1 = 1/2$). Nine 2D profiles with equally-spaced targeted gain levels starting from 0.7 dB and ending with 6.3 dB (slightly higher than 60\% merit of Raman pumping as proposed in \cite{7254112}) are generated by applying the DE to solve Eq. \ref{eq:so}. In Fig. \ref{subplot_simulation_flat_v1}(a) the power evolution of the designed 2D profiles with their corresponding flat target gains $\textbf{g}^t$ are shown. The corresponding target gain level and the designed spectral gain levels of each 2D profile is demonstrated in Fig. \ref{subplot_simulation_flat_v1}(b). Moreover, the average cost over the nine 2D profiles for each DE iteration is calculated and shown in Fig.\ref{subplot_simulation_flat_v1} (c), which asserts that no considerable accuracy improvement after 260 iterations is obtained.
Additionally, Fig. \ref{subplot_simulation_flat_v1}(d) shows that the resulting cost values are less than 0.6 dB  for target gains levels less than 5.6 dB. For all cases with different target gain levels, the obtained pump power values are shown in Fig.\ref{Optim_pump_power_values_subplot_simulation}. It is illustrated that for the case with the highest target gain level (6.3 dB), the power value of the pump $p_4$ reaches its upper-bound 23 dB value and does not allow the DE to improve further. This results confirm that in case the amplifier setup is provided with pumps with higher upper-bound power values, high target gains can be achieved accurately using the DE framework.

% \begin{figure}[h!]
% \centering\includegraphics[width=8cm]{CNN_assisted_DE_max_error_new1.pdf}
% \caption{CNN assisted DE performance.}
% \label{CNN_assisted_DE_max_err}
% \end{figure}

\label{sec:bibtex}

\section{Conclusion}
The DE and the CNN-assisted DE frameworks are experimentally validated for designing 2D power evolution profiles using Raman amplifiers. For test 2D profiles, the CNN model achieves less than 0.4 dB MAE on average while it is inaccurate for 2.2\% of the 2D profiles in the test data-set. Addressing these profiles, CNN-assisted DE is applied to fine-tune the pump powers values, showing more than 1 dB improvement in average over the CNN results. In a second designing scenario, the DE framework is employed to design 2D profiles with two different objectives: 1) flat gain levels and, 2) minimum spectral power deviation. The results assert that the proposed frameworks can be effectively used to design 2D profiles, by online tuning of the pump power values in an amplifier setup under test.

\begin{backmatter}
\bmsection{Funding}European Research Council (ERC-CoG FRECOM grant no. 771878); The Villum Foundation (OPTIC-AI grant no. 29334); The Italian Ministry for University and Research (PRIN 2017, project FIRST).

\bmsection{Disclosures}The authors declare no conflicts of interest.
\bmsection{Data Availability Statement}Data underlying the results presented in this paper are not publicly available at this time but may
be obtained from the authors upon reasonable request.

\end{backmatter}
%%%%%%%%%%%%%%%%%%%%%%% References %%%%%%%%%%%%%%%%%%%%%%%%%

%%%%%%%%%% If using BibTeX:
\bibliography{sample}

%%%%%%%%%% If preparing manually:
% \begin{thebibliography}{1}
% \newcommand{\enquote}[1]{``#1''}

% \bibitem{Zhang:14}
% Y.~Zhang, S.~Qiao, L.~Sun, Q.~W. Shi, W.~Huang, L.~Li, and Z.~Yang,
%   \enquote{Photoinduced active terahertz metamaterials with nanostructured
%   vanadium dioxide film deposited by sol-gel method,}
%   {\protect\JournalTitle{Optics Express}} \textbf{22}, 11070--11078 (2014).

% \bibitem{OSA}
% {Optical Society}, \enquote{{OSA Publishing},}
%   \url{http://www.osapublishing.org}.

% \bibitem{FORSTER2007}
% P.~Forster, V.~Ramaswamy, P.~Artaxo, T.~Bernsten, R.~Betts, D.~Fahey,
%   J.~Haywood, J.~Lean, D.~Lowe, G.~Myhre, J.~Nganga, R.~Prinn, G.~Raga,
%   M.~Schulz, and R.~V. Dorland, \enquote{Changes in atmospheric consituents and
%   in radiative forcing,} in \enquote{Climate Change 2007: The Physical Science
%   Basis. Contribution of Working Group 1 to the Fourth assesment report of
%   Intergovernmental Panel on Climate Change,}  S.~Solomon, D.~Qin, M.~Manning,
%   Z.~Chen, M.~Marquis, K.~B. Averyt, M.~Tignor, and H.~L. Miler, eds.
%   (Cambridge University Press, 2007).

% \end{thebibliography}

\end{document}